# Steady State Multiplicity in a Polymer Electrolyte Membrane Fuel Cell

Ee-Sunn J. Chia, Jay B. Benziger, Ioannis G. Kevrekidis

Department of Chemical Engineering, Princeton University, Princeton, NJ 08544

**Introduction.** Polymer electrolyte membrane fuel cells (PEM-FC) have received widespread attention as an alternative power source (EG&G Services, 2000). Fuel cells for mobile applications will necessarily be non-steady-state (variable loads in power plants, climate effects on temperature and humidity, but also fast changes, such as vehicle acceleration). Predictive models of fuel cell *dynamic* performance that correctly incorporate the interplay of reaction and transport processes, are therefore critical. A differential PEM fuel cell reactor was constructed to examine kinetics and dynamics; this reactor design, which bypasses the additional complexity of two- and three-dimensional integral reactors (Springer et al., 1991; Bernardi and Verbrugge, 1992; Fuller and Newman, 1993; Janssen, 2001), focuses on reaction/transport dynamic coupling under well-defined conditions. Ignition/extinction phenomena and multiple steady states were reproducibly demonstrated by Benziger et al. (2003). These experiments demonstrate that the relation of proton transport with water activity in the PEM membrane underpins the observed dynamical phenomena. Water ionizes and shields stationary anions in the membrane, which causes proton transport to increase by orders of magnitude.

Here we present and analyze a simplified model that embodies what we believe to be the essential physics controlling PEM fuel cell dynamics. A remarkable analogy exists between water management in the differential PEM-FC and energy balance in the classical exothermic stirred tank reactor. Water, the reaction product in the PEM-FC autocatalytically accelerates the reaction rate by enhancing proton transport through the PEM. This is analogous to the Arrhenius temperature-based rate acceleration due to the heat produced by an exothermic reaction. We modify the established textbook analysis of heat autocatalyticity in a CSTR (van Heerden, 1953; Aris, 1965; Perlmutter, 1972; Uppal et al., 1974; Schmitz, 1975) to present water management autocatalyticity in a stirred tank reactor PEM-FC. Steady states arise at the intersection of a (linear) water removal curve and a (sigmoidal) water production curve. Having established the autocatalyticity analogy between the exothermic CSTR (Aris, 1965) and what we will now call the STR-PEM, we can use modeling tools for the dynamic/parametric analysis of chemical reactors (continuation, singularity theory, numerical stability and bifurcation analysis) to explore the STR-PEM dynamic and parametric operation (Doedel, 1981; Farr and Aris, 1986; Balakotaiah and Luss, 1982a,b).

**The STR-PEM and its model.** A schematic of the hydrogen-oxygen STR-PEM fuel cell is shown in Figure 1a. Hydrogen and oxygen gases are fed into the anode and cathode chambers at inlet molar rates of $n_{H_2}^{in}$ and $n_{O_2}^{in}$ (mol/s) respectively. Hydrogen molecules are dissociatively adsorbed at the anode and oxidized to protons. Electrons travel through an external load resistance $R_L$, (Ω). Protons diffuse through the PEM under an electrochemical gradient to the cathode. Oxygen molecules adsorbed at the cathode, are reduced by the electrons and react with the protons to produce water. In this paper we consider the autohumidified fuel cell; no water vapor is contained in the feed gases. The membrane is humidified by the water produced. The anode and cathode chambers are modeled as two stirred tank reactors sandwiching the PEM. Figure 1b is an electrical circuit representation of the fuel cell model where $R_M$ (Ω) denotes the membrane resistance to proton transport and $V$ (Volts) is the voltage associated with the chemical potential difference across the membrane. The operating parameters are the fuel cell temperature $T$ (K), external load resistance $R_L$, and inlet molar rates $n_{H_2}^{in}$ and $n_{O_2}^{in}$. The rate of water production in the fuel cell is expressed through the cell current $i$ (Amps).

The polymer electrolyte membrane serves as the medium for proton transport. We use a Nafion™ membrane, which is a partially substituted perfluoro-sulfonic acid ionomer. Water absorbed into the membrane ionizes the sulfonic acid groups, and facilitates proton transport via a hopping mechanism between fixed ions. Membrane humidification is essential for PEM-FC operation. Proton conductivity increases exponentially with membrane water content (Hsu and Gierke, 1982; Eikerling et al., 1997; Thampan et al., 2000; Paddison, 2001; Yang, 2003).

Our model is centered on the water balance in the cell, expressed in terms of a single dynamic variable: the dimensionless water activity in the membrane, $a_w$. At this level of modeling, this activity is assumed uniform over the membrane. This activity is defined as the gas partial pressure of water $P_w$, at equilibrium with the membrane divided by the vapor pressure of water at the cell temperature $P_w^o(T)$. This single dynamic variable determines the cell dynamics through the following connections:

a) The water content of the membrane, $\lambda(a_w)$, expressed in terms of water molecules per membrane sulfonic acid group, has been correlated experimentally by Yang as (2003):

$$\lambda = 14.9 a_w - 44.7 a_w^2 + 70.0 a_w^3 - 26.5 a_w^4 - 0.446 a_w^5 \qquad \frac{[H_2O]}{[SO_3^-]} \qquad (1)$$

b) The membrane resistance to proton transport, $R_M(a_w)$. This has been found experimentally to have a negligible temperature dependence, and for a membrane of thickness $l$ cm and area $A$ cm$^2$ has been correlated as:

$$R_M(a_w) = 10^7 \exp(-14 a_w^{0.2}) \frac{l}{A} \qquad [\Omega] \qquad (2)$$

c) Through the membrane resistance, $a_w$ also determines the cell reaction rate (current, $i$). Indeed, the cell potential $V$ is almost independent of reactant concentration as long as the reactant utilization is less than ~80%. We can therefore (Figure 1b) write

$$i = \frac{V}{R_M(a_w) + R_L} \qquad [A] \qquad (3)$$

We neglect kinetic limitations for the transport of water between the membrane and the gas phase in the anode and cathode chambers, and assume that the water vapor pressure $P_w$ is at equilibrium with the membrane water activity. The partial pressure of water in the two chambers is the product of the water vapor pressure at the cell temperature and the membrane water activity $P_w = P_w^o(T) a_w$. The cell mass balance for water is shown below, where $\mathcal{F}$ is Faraday's constant (96,500 C/mol), while $V_A^g$ and $V_C^g$ (L) represent the volume of gas in the anode and cathode chamber respectively.

$$\left[ N_{SO_3} \frac{d\lambda(a_w)}{da_w} + (V_A^g + V_C^g) \frac{P_w^o}{RT} \right] \frac{da_w}{dt} = 0.5 \frac{i(a_w)}{\mathcal{F}} - (F_A + F_C) P_w^o \frac{a_w}{RT} \quad (4)$$

$$\underbrace{\phantom{xxxxx}}_{\text{water in membrane}} \quad \underbrace{\phantom{xxxxx}}_{\text{water in gas phase}} \quad \underbrace{\phantom{xxxxx}}_{\text{water production}} \quad \underbrace{\phantom{xxxxx}}_{\text{water removal}}$$

Clearly, mass transfer coefficients will be required at a subsequent modeling level to quantify the mass transfer between the membrane and the gas phase. We assume constant pressure operation ($P_{tot}$ =1 bar, same in both chambers). In a further simplification, one may consider that the reactants enter the cell at dilute concentrations, so that the molar flow rate can be considered constant. We will also consider the case where the reactant feeds are pure components; the exit flow rates $F_A$ and $F_C$ in the above equation are then defined by the inlet molar flow rates, $n_{H_2}^{in}$ for the anode and $n_{O_2}^{in}$ for the cathode, the operating pressure $P_{tot}$ and the reaction. Given our equilibrium assumption for water, which equipartitions the water produced in the two chambers, $F_C$ is independent of the water activity, while, for the anode, $F_A$ will now depend on $a_w$.

$$F_C = \frac{n_{O_2}^{in} RT}{P_{tot}}, \quad F_A = \frac{RT}{P_{tot}} \left[ n_{H_2}^{in} - 0.25 \frac{i(a_w)}{\mathcal{F}} \right] \quad [L/s] \quad (5)$$

Equation 4, either with $F_A$ and $F_C$ constant, or alternatively given by Eq. 5 above is our working model. For pure feed operation the gas phase pressure variation of hydrogen (anode) and oxygen (cathode) directly follow:

$$\frac{dP_{H_2}}{dt} = \frac{RT}{V_g^A} \left[ n_{H_2}^{in} - F_A \frac{P_{H_2}}{RT} - 0.5 \frac{i(a_w)}{\mathcal{F}} \right], \quad \frac{dP_{O_2}}{dt} = \frac{RT}{V_g^C} \left[ n_{O_2}^{in} - F_C \frac{P_{O_2}}{RT} - 0.25 \frac{i(a_w)}{\mathcal{F}} \right] \quad (6)$$

To illustrate the water balance we graphically show, in Figure 2a the water production and water removal curves (the first and second terms on the right-hand-side of Eq. 4 respectively). For simplicity, we first consider the dilute reactant feed case ($F_A$, $F_C$ constant). The water removal curve is a straight line in this plot, while the experimentally correlated water production curve is sigmoidal, due to the exponential enhancement of proton transport with membrane water activity. Steady states correspond to the intersection of the two curves; stable (unstable) ones are marked with filled (empty) circles. Slight perturbations around the middle (unstable) steady state in Figure 2a will, in a fashion analogous to the autothermal reactor stability arguments, drive the STR-PEM towards the upper or lower stable steady states.

In pure feed, constant pressure operation, the anode and cathode outlet gas flow rates $F_A$ and $F_C$ are not constant, and this causes changes in the curvature of the water removal curve. The qualitative sense of the results and their dependence on operating parameters provided by the diagram, however, still holds. Figure 2c shows water production predicted by the following modified expression for the membrane resistance:

$$R_M(a_w) = \left\{ 10^7 \exp(-14 a_w^{0.2}) - \left[ y_o + \frac{\mu}{w\sqrt{\pi/2}} \exp\left[ -2\left( \frac{a_w - x_c}{w} \right)^2 \right] \right] \right\} \frac{l}{A} \quad (7)$$

This expression (with $\mu = 76$, $y_o = 0.01$, $w = 0.07$, $x_c = 0.08$) attempts to capture the experimental observation of "pinning" of the membrane resistance over certain intermediate ranges of water activity. This pinning, which causes the membrane resistance to vary in a stepwise fashion arises, we believe, from the mechanical constraints the electrode assembly imposes on the swelling membrane. As Figure 2c shows, this can lead to five total (three stable) steady states, consistent with experimental observations (Benziger et al., 2003).

The water balance diagrams in Figure 2 can be used to explore the dependence of the dynamics on operating parameters. Temperature mainly affects water removal through the water vapor pressure; higher temperatures increase the slope of the water removal curve. The load resistance, on the other hand alters water production: increasing $R_L$ depresses the water production curve, especially at high water activities. Higher reactant flow rates tilt the water removal curve higher. It is interesting to observe that prehumidification of the reactant feeds translates the water removal line/curve to the right. These considerations elucidate the way steady state multiplicity depends on parameters. Before a more systematic presentation of these results, a caveat: the validity of both the data and the model predictions is limited to water activities less than (possibly approaching) one. Predictions beyond this limit are invalid; they signal conditions under which condensation of liquid water will occur (something experimentally observed). An extension to a two-phase model will be required to quantify such operation.

**Computational Results.** We use continuation/numerical bifurcation techniques implemented in the software known as AUTO to identify regions in parameter space where steady state multiplicity exists (Doedel, 1981). Four experimentally controllable parameters, $T$, $R_L$, $n_{H_2}^{in}$, and $n_{O_2}^{in}$ are varied in our computations. Since we assume that water in the membrane and vapor phase are in equilibrium, constant pressure operation at 1 bar indirectly establishes a maximum temperature (373 K) allowable in the fuel cell.

The results are organized in terms of three two-parameter bifurcation diagrams ($T$ and $R_L$ in Figure 3; $T$ and $n_{H_2}^{in}$ in Figure 4a; $n_{O_2}^{in}$ and $n_{H_2}^{in}$ in Figure 4b). Regions of one, three and five steady state behavior are marked I, III and V respectively. Blowups of the first two diagrams are included as insets, to provide a better feeling of steady state multiplicity over larger parameter ranges. Representative one-parameter cuts, marked on the two-parameter diagrams, are included, to illustrate the various ignition-extinction transitions involved. The ordinate of the one-parameter diagrams show the steady state cell current density, reflecting water production in the fuel cell; "ignited" ("extinguished") branches are characterized by high (low) current densities. Solid curves in the one-parameter bifurcation diagrams denote stable steady states while dashed curves correspond to unstable ones.

Figure 3 depicts a two-parameter bifurcation diagram in the load resistance - temperature plane with inlet gas rates of 5.7 x $10^{-6}$ mol/s (10mL/min). Figures 3($\alpha_1$)-($\alpha_3$) constitute representative one-parameter diagrams with respect to temperature at load resistance values of 15 Ω, 2.5 Ω, and 1.0 Ω respectively. For all these resistances, going below a certain temperature (~325 K) will cause an ignition; this is because the vapor pressure of water is so low, that water is not convected away and builds up in the membrane. For the large load resistance of 15 Ω ($\alpha_1$), a maximum of three steady states is attainable. In addition, the cell current density at the ignited state is significantly lower compared to current densities obtained for smaller load resistances.

Similar one-parameter continuation diagrams in the load resistance at different temperatures are shown in Figures 3($\beta_1$)-($\beta_3$) for temperatures of 322 K, 350 K, and 356 K respectively. As one might expect, above a certain critical load resistance, the cell extinguishes.

A two-parameter continuation of turning points in the hydrogen feed and temperature plane is shown in Figure 4a. Note that the dotted curve in Figure 4a has been shifted a little to the right to increase visibility of the two smaller three steady state regions. Representative one-parameter continuation diagrams with respect to temperature are shown at constant hydrogen feed rates of 5.7 x $10^{-6}$ mol/s ($\alpha_1$) and 4.0 x $10^{-5}$ mol/s ($\alpha_2$). Corresponding one-parameter diagrams with respect to hydrogen feed are depicted for fixed temperatures of 330 K ($\beta_1$) and 353 K ($\beta_2$). Increasing the hydrogen feed convects the water product out faster, and dries the membrane out; at higher temperatures the vapor pressure of water increases, and the "final extinction" occurs at lower hydrogen flow rates. In the one-parameter continuations with respect to temperature (Figures 4($\alpha_1$)-($\alpha_2$)) the most noticeable feature is the proximity of the two ignition points apparent in each. Cusp points and double-limit points (suggestive of a swallowtail organizing center) appear in our study (Guckenheimer and Holmes, 1983). No codimension two bifurcations were visible in the two-parameter diagram in the inlet hydrogen and inlet oxygen feed rate planes in the region covered in Figure 4b.

**Discussion.** We have demonstrated, through a simplified but physically reasonable model of an STR-PEM fuel cell steady state multiplicity, caused by the autocatalytic nature of the interplay between water (a product of the reaction) and the reaction rate, which is enhanced through membrane humidification. This creates a succinct analogy with the energy balance of an autocatalytic CSTR: water activity in the STR-PEM corresponds to temperature in the CSTR. Water production and removal are analogous to heat production and removal, and the corresponding straight and sigmoidal curve shapes remarkably persist in the analogy. Inlet feed humidification (STR-PEM) is analogous to inlet feed temperature (CSTR); load resistance (STR-PEM) to reaction enthalpy (CSTR); finally, feed flow rates convect away the reaction product in both cases. Membrane swelling against mechanical constraints adds a five steady state "twist" to the STR-PEM dynamics, reminiscent of the richness of dynamics of exothermic reactions in series in a CSTR.

How much further can this analogy be taken? We already have preliminary experimental results of sustained periodic and even chaotic oscillations in the differential STR-PEM. In an integral PEM-FC reactor there is evidence of high-current, high local water content "wet spots," clearly analogous to tubular reactor "hot spots." Reported irreproducibilities in fuel cell stacks may well be explained by secondary ignitions due to swelling phenomena. Effective control for fuel cell vehicles will not be possible until the physical mechanisms and time scales of these complicated dynamics are understood.

**Acknowledgements**. The authors acknowledge support through the AFOSR (Dynamics and Control).

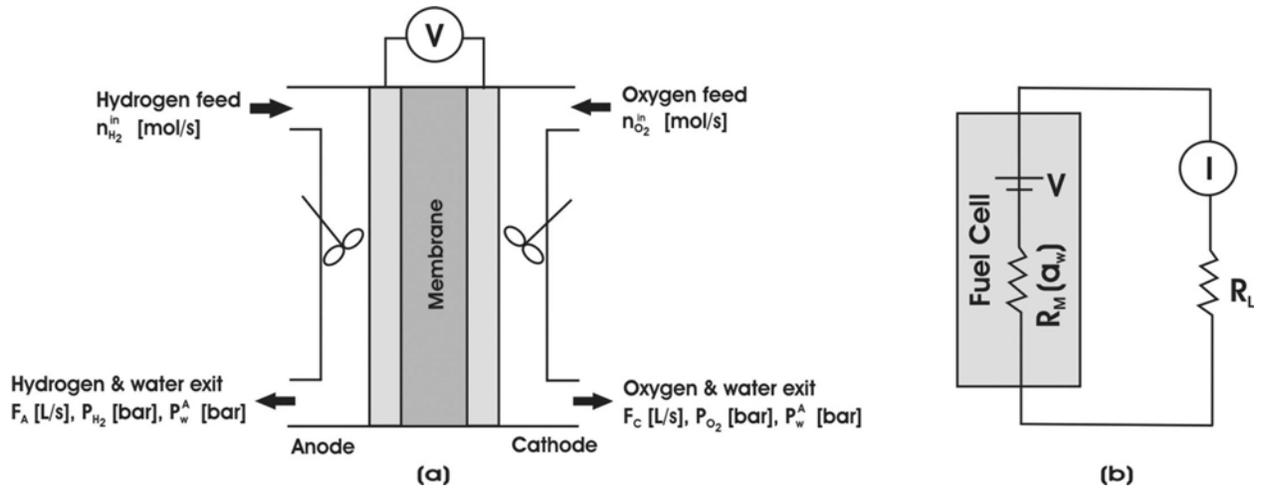

Figure 1: PEM fuel cell model
(a) The anode and cathode chambers in the PEM fuel cell are modeled as two stirred tank reactors in series. Non-humidified hydrogen and oxygen are fed into the reactor while the unreacted gases and product water exit from each chamber. (b) An electrical circuit equivalent of the PEM fuel cell model.

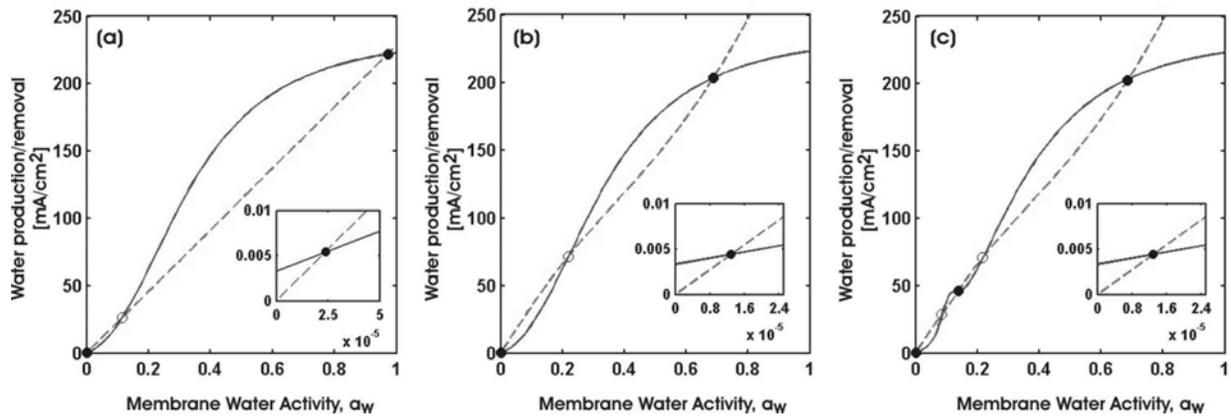

Figure 2: Water management and steady state multiplicity in a PEM fuel cell.
Water production (solid curve) and removal (dashed curve) are illustrated as functions of the membrane water activity $a_w$. The filled (empty) circles denote stable (unstable) steady states.
(a) Dilute reactant feed [$T = 298$ K, $R_L = 1.2$ $\Omega$, $F_A = F_C = 1.42 \times 10^{-3}$ L/s (85 mL/min)]
(b)-(c) Pure reactant feed [$T = 353$ K, $R_L = 1.2$ $\Omega$, $P_T = 1$ bar, $n_{H2}^{in} = n_{O2}^{in} = 5.7 \times 10^{-6}$ mol/s].

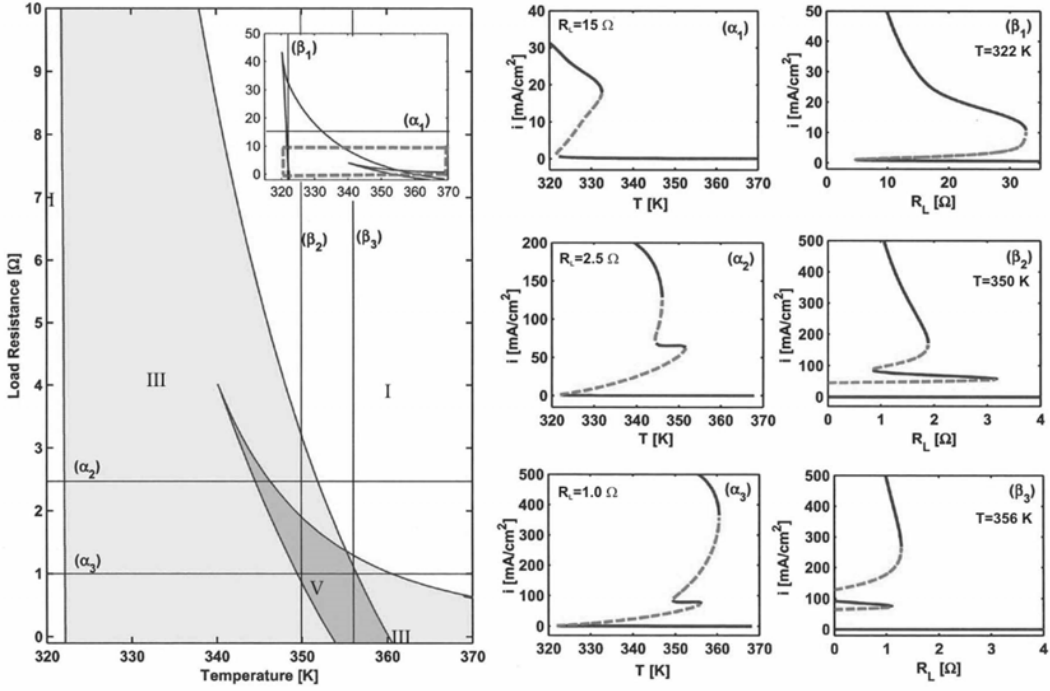

Figure 3: Two-parameter bifurcation diagram in $T$ and $R_L$ for $n_{H2}^{in} = n_{O2}^{in} = 5.7 \times 10^{-6}$ mol/s. Regions of one, three, and five steady states are marked by I, III, and V. Corresponding one-parameter continuation diagrams with respect to temperature and to load resistance are shown in $(\alpha_1)$-$(\alpha_3)$ and $(\beta_1)$-$(\beta_3)$ respectively. The solid (dashed) curves denote stable (unstable) steady states.

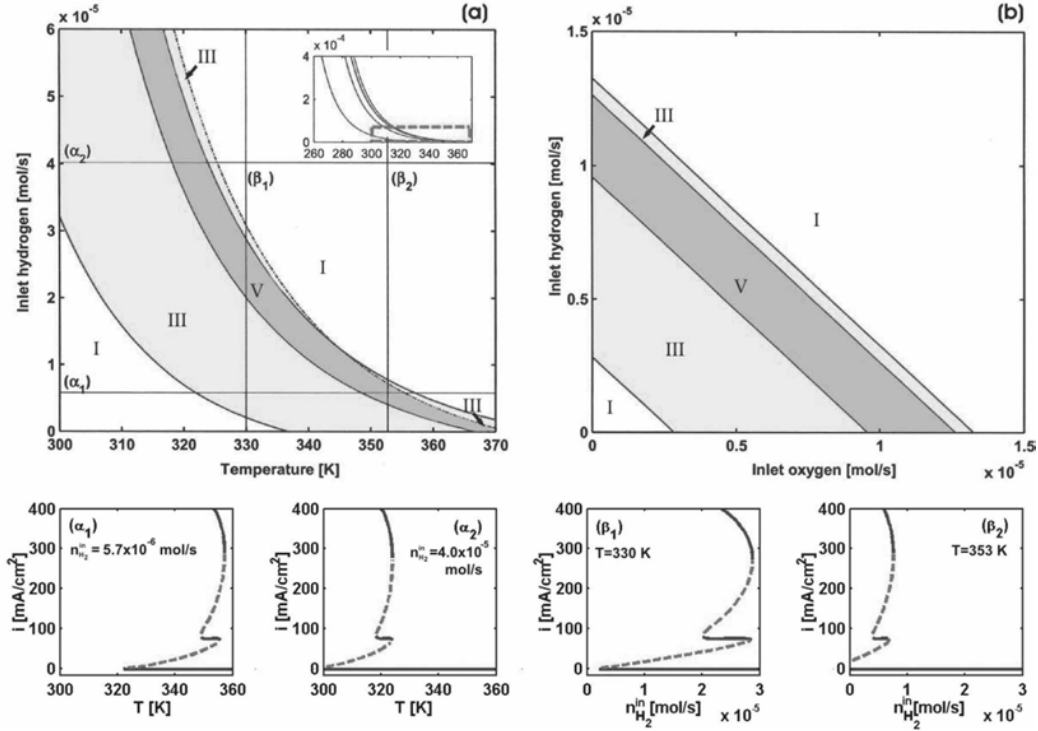

Figure 4: Two-parameter bifurcation diagram in
(a) $T$ and $n_{H2}^{in}$ for $R_L=1.2\ \Omega$, $n_{O2}^{in} = 5.7 \times 10^{-6}$ mol/s
(b) $n_{O2}^{in}$ and $n_{H2}^{in}$ for $T=353$ K, $R_L=1.2\ \Omega$
Corresponding one-parameter continuation with respect to temperature and to hydrogen feed are shown in $(\alpha_1)$-$(\alpha_2)$ and $(\beta_1)$-$(\beta_2)$ respectively.